\documentclass[11pt]{article}
\usepackage{fancyhdr}
\def\eqalign#1{\null\,\vcenter{\openup\jot
  \ialign{\strut\hfil$\displaystyle{##}$&$\displaystyle{{}##}$\hfil
      \crcr#1\crcr}}\,}

\usepackage{color}
\usepackage{hyperref}
\def\iniz{\setcounter{equation}{0}{%
\rhead{\thepage}\lhead{{{{\small\bf\thesection:}
\small \ \SEC\ \  \tiny\today}}}}}

\let\a=\alpha \let\b=\beta     \let\e=\varepsilon
  \let\h=\eta   \let\th=\vartheta  \let\l=\lambda
\let\m=\mu    \let\n=\nu         \let\p=\pi    
\let\s=\sigma    \let\f=\varphi 
\let\ch=\chi     
\let\G=\Gamma \let\D=\Delta  \let\L=\Lambda

\def\V#1{{\bf#1}}\def\rhs{{\it r.h.s.}\ }
\def\*{\vskip 3mm}\def\0{\noindent}
\def\be{\begin{equation}}
\def\ee{\end{equation}}
\def\bea{\begin{eqnarray}}
\def\eea{\end{eqnarray}}

\def\LL{{\mathcal L}}

\def\QQ{{\mathcal Q}}
\let\dpr=\partial\let\fra=\frac
\def\ie{{\it i.e.}\ }
\def\eg{{\it e.g.}\ }
\def\Eq#1{\label{#1}}
\def\equ#1{(\ref{#1})}
\def\lis#1{\overline{#1}}
\def\defi{{\buildrel def\over=}}
\def\media#1{{\Blangle\,#1\,\Brangle}}

\def\otto{\,{\kern-1.truept\leftarrow\kern-5.truept\to\kern-1.truept}\,}
\def\wt#1{{\widetilde#1}}

\def\tende#1{\,\vtop{\ialign{##\crcr\rightarrowfill\crcr
 \noalign{\kern-1pt\nointerlineskip} \hskip3.pt${\scriptstyle
 #1}$\hskip3.pt\crcr}}\,}
\def\ie{{\it i.e.\ }}

\newdimen\xshift \newdimen\xwidth \newdimen\yshift \newdimen\ywidth
\def\ins#1#2#3{\vbox to0pt{\kern-#2pt\hbox{\kern#1pt #3}\vss}\nointerlineskip}

\def\eqfig#1#2#3#4#5{
\par\xwidth=#1pt \xshift=\hsize \advance\xshift
by-\xwidth \divide\xshift by 2
\yshift=#2pt \divide\yshift by 2%
{\hglue\xshift \vbox to #2pt{\vfil
#3 \includegraphics{#4.eps}
}\hfill\raise\yshift\hbox{#5}}}

\def\Brangle {{\mbox{\boldmath$ \rangle$}}}
\def\Blangle {{\mbox{\boldmath$ \langle$}}}

\begin{document}

\kern-1cm
{\color{red}
\centerline{\Large\bf Renormalization group and divergences 
}} \*

{\color{blue}%
\centerline{\tt Giovanni Gallavotti}} 
{\color{blue}\centerline{\small INFN-Roma1 and Rutgers
  University}} {\color{red}\centerline{\tiny\today} }

\* \0{\bf Abstract: \it Application of asymptotic freedom to the
  ultraviolet stability in Euclidean quantum field theories is revisited
  and illustrated through the hierarchical model making also use of a few
  technical developments that followed the original works of Wilson on the
  renormalization group. 
}

\*
\0{Key words: \it Renormalization group, Coupling renormalization,
Hierarchical model, Quantum field theory}

\* 
\def\SEC{Euclidean quantum fields}
\section{\SEC}\iniz\label{sec1}

The first examples that Wilson worked out about the constructions of the
renormalization flow are in two quite similar works, \cite{Wi965,Wi970}.
The second is slightly simpler because it deals with a system consisting
entirely of spins (\ie described by bounded operators). It essentially
introduces the renormalization group method via the consideration of
hierarchical models as a tool to understand the essence of renormalization
theory. The hierarchical models also appeared explicitly essentially
at the same time in the work of \cite{Dy969} devoted to the theory of phase
transitions (in $1$ dimension) but not in relation to renormalization
theory: the intimate relation between the two domains (statistical
mechanics and quantum field theory) was a consequent development.

The work \cite{Wi970} ideally foreshadows the theory of the Kondo effect
developed shortly afterwards and presented in detail in \cite{Wi975};
the papers \cite{Wi971a,Wi971b} reduce to the theory of a dynamical system
the study of the critical point in the Ising model: a breakthrough making
possible, for the first time, a computer aided approach to the calculation
of critical exponents in dimension $<4$. At the same time it gave a
solution, via the same recursion, to the ultraviolet stability in QFT of
dimension $<4$, a classical renormalization problem studied until then by
rather different methods, \cite{Ne966,GJ981}.

The work of Wilson that most influenced {\it constructive theory} of
quantum fields has been the analysis of the hierarchical model performed
applying his view of renormalization, \cite{Wi971a},\cite[Eq.(23)]{Wi972}
to scalar field theory: it made crystal clear that the divergences removal
(already known since the early days of renormalization theory to be a
``multiscale problem'', \cite{He969}) was reducible to controlling a
dynamical system governing the evolution, as the ``scale'' changed up from
the ultraviolet (from short distances to distances of $O(1)$) or up in the
infrared (from distances of $O(1)$ to large distances), of a few ``running
couplings'' with a technique that unified conceptually the quantum field
theory renormalization and the classical critical point theory via the new
concept of asymptotic freedom, to the emergence of which his work gave an
important contribution, \cite{Gr999,Ho999b}.

The hierarchical model analysis for the scalar $\l\f^4$ field in space-time
dimensions $\le3$, performed following Wilson's renormalization methods,
teaches how to treat functional integrals (at least in the asymptotically
free theories) as chains of ``naive'' sums. In the end it shows, for
instance in scalar QFT at low dimension, that there is no divergence
problem if the analysis is properly set up: because the physically
interesting quantities (like the ``Schwinger functions'') are expressed as
power series in the running couplings with no divergences at all. 

This is an important result, although the model is a simplified version of
a theory, the ``$\l\f^4$ field theory'', which at the time ``had no obvious
application anywhere in elementary particle physics'', \cite{Ho999b}.

Divergences arise if the running couplings are expanded in power series of
the constants in the Lagrangian function; the point being the lack of
analyticity of the running couplings in terms of the parameters present in
the Lagrangian, called ``bare constants''. Attempting an ({\it
  unnecessary}) expansion of the running couplings in terms of the
parameters present in the Lagrangian, called bare constants, results in
divergent expressions.

In Wilson's approach bare constants will never appear (and therefore the
accompanying divergences will never arise): the theory will be described by
the sequence of the running couplings which are related to their values on
the physical scale \footnote{\small The observer's length and time scales
  are by definition of $O(1)$.} by a map, called the beta function. At
least not in theories which are asymptotically free: in the others, which
represent many physically relevant problems, like the critical point
theory, the question is still very hard as it relies on the possible
existence of non trivial fixed points for the map describing the running
constants flow through the different scales.

Implicitly the hierarchical model was introduced already in \cite{Wi970}
(related to ``meson theory'') and it was preceded by an even simpler
version (related to the ``Lee model'') \cite{Wi965}. In its simplest
version it is a model for the Euclidean $\f^4$-theory in the ultraviolet
region. This is a theory which in space-time dimension $\le3$ is
asymptotically free in the ultraviolet region and asymptotically non trivial
in the infrared region: and the basic mathematical problem is to give a
meaning to the functional integral
\be Z=\int e^{-\int_\L(\l\f_x^4+\m \f_x^2+\n)dx} \Big[C
\,e^{-\frac12\int((\dpr _x\f_x)^2 +\f_x^2)dx}\prod_xd\f_x\Big]
\Eq{e1.1}\ee
where $\f_x$ is a function on $\L$ and, in the ``ultraviolet problem'', the
integral in the exponent is over a finite volume $\L$, \eg a cube for
simplicity (if $d=3$) or a square (if $d=2$).  The easy case $\l=\m=\n=0$,
``free field'', corresponds to interpreting the quantity in square
brackets in Eq.\equ{e1.1} as a Gaussian probability
distribution assigning average value $\media{\f_x\f_y}$ to the product
$\f_x\f_y$ as:
\be\G(x,y)\defi\media{\f_x\f_y}=\frac1{(2\p)^d}\int \frac{e^{i p(x-y)}}{1+p^2}
d^dp\simeq const \frac{e^{-|x-y|}}{|x-y|^{d-2}}
\Eq{e1.2}\ee
which, through the rules for Gaussian integrals (``Wick's rules''), defines
all the averages $\media{\f_{x_1}\f_{x_2}\cdots\f_{x_{2n}}}$.

The basic difficulties can be seen from the fact that if $d\ge2$ then
$\media{\f_x^2}=+\infty$: with the consequent failure of any attempt to
evaluate $Z$ through an expansion in powers of
$\l,\m,\n$, for instance, the integral in Eq.\equ{e1.1} or
\be\frac{1}Z\int \f_x\f_y\,e^{-\int_\L(\l\f_x^4+\m
  \f_x^2+\n)dx} 
\Big[C e^{-\frac12\int((\dpr _x\f_x)^2+\f_x^2)dx}\prod_xd\f_x\Big]
\Eq{e1.3}\ee
Yet it is well known that the founding fathers devised a resummation
scheme, the ``renormalization'', of the series so that divergences would 
disappear.

In the work \cite{Wi965} Wilson undertook to define an algorithm that would
produce the resummation of the formal series (with divergent coefficients)
transforming it into a power series of a new sequence of {\it finite}
constants related to each other as subsequent elements of a trajectory of a
map in a finite dimensional space ({\it very low dimensional}, actually one
dimensional in the quoted paper) with initial data suitably restricted.

The simple but new idea was that the functional integral had to be thought of
as a sequence of almost identical integrals each of which simple enough to
be computable naively. The hierarchical model realizes a paradigmatic
case.

\* \def\SEC{The hierarchical model}
\section{\SEC}\iniz\label{sec2}

Before discussing in detail the model it is interesting to quote
what appears to be its birth moment:

\*\0''{\it In this approximation the free-meson field has been replaced by
  independent harmonic oscillators for each phase space cell, with a
  frequency depending only on the mean momentum of the cell. The
  interaction of the meson field with the source has been replaced by an
  interaction of those oscillators located at the origin (where the source
  is) with the source. The remaining terms of the original Hamiltonian are
  to be considered as a perturbation}'' \cite[p.455]{Wi965}. \*

As will be seen below this viewpoint, very clearly presented again in
\cite[p.3184]{Wi971b}, where the following heuristic remark summarizes
another key idea:
\*

\0{\it 
This means that $s_L(\V x)$ does not vary enormously
within a block of size $L$ and for qualitative purposes
one can think of $s_L(\V x)$ within a block as if it were
a single block variable},
\*

\0and in \cite[Eqs.(23),(33)]{Wi972}, opens the way to a totally new
conception of renormalization theory through functional integrals: I allows
himself to remember here a talk by Wilson at the University of Roma in the
early '70's. There I was amazed to see the way and ease he was using to
compute functional integrals: it was in sharp contrast to what I was used
to after learning the mathematical theory of Brownian motion (no functional
spaces in sight, no Banach spaces, no subtle almost everywhere statements,
...), and the procedure seemed to me far from mathematical rigor. I raised
hand and signified my disappointment: the lapidary reply was just ``you do
not understand functional integration''. {\it Therefore} I tried to
understand why and shortly afterwards I was working intensely on the
renormalization group in scalar quantum fields, using the methods that he
had described, and I kept doing so for the next two decades.  \*

Imagine $\L$ of side $L$ and paved by cubes or squares $\D$ of side
$2^{-n}L$, $n=0,1,\ldots$; the pavements $\QQ_n$ will be said to have ``scale
$n$''. To each $\D$ associate a normal Gaussian random variable $z_\D$ with
distribution $P(d z_\D)$ and define
\be \f_x\,\defi\,\sum_{n=0}^\infty \sum_{x\in \D\in \QQ_n} 2^{\frac{d-2}2n}
z_\D,\qquad\  P(dz_\D)\,\defi\,\frac{e^{-\frac12z_\D^2}}{\sqrt{2\p}}dz_\D
\Eq{e2.1}\ee
 The distribution of the $\f_x$'s thus constructed is ``quite close'' to the
 Gaussian process defined by Eq.\equ{e1.2}. Let $d_h(x,y)$ denote
 $2^{-n(x,y)}$ with $n(x,y)-1$ being the scale of the smallest $\D$ that
 contains both $x$ and $y$; then $d_h(x,y)$, called {\it dyadic distance of
   $x,y$}, will often enough be close to the actual distance between $x,y$:
 in the sense that the average $\media{\f_x\f_y}$ of the product of two
 $\f$'s as defined by Eq.\equ{e2.1} is
\be C(x,y)\defi\media{\f_x\f_y}=\cases{
-\log_2{d_h(x,y)}& if $d=2$\cr
\frac1{d_h(x,y)^{d-2}}
\frac{1- d_h(x,y)^{d-2}}{2^{d-2}-1}\simeq \frac1{d_h(x,y)^{d-2}}& if $d>2$}
\Eq{e2.2}\ee
Certainly the value of the field $\f_x$ is infinite for every $x$:
nevertheless $C(x,y)<\infty$ if $x\ne y$.  A precise meaning of
Eq.\equ{e1.1},\equ{e1.2} can be defined via a ``regularization procedure'':
define $\f_x^{[\le N]}$ as
\be \f_x^{[\le N]}\defi\sum_{n=0}^{N} 
\sum_{x\in \D\in \QQ_n} 2^{\frac{d-2}2n}
z_\D\Eq{e2.3}\ee
which is a well defined finite sum and therefore

\be Z_N\defi \int e^{-\int_\L(\l(\f^{[\le N]}_x)^4+\m
  (\f^{[\le N]}_x)^2+\n)dx}  P(d\f)\Eq{e2.4}\ee
is well defined if $P(d\f)=\prod_\D P(d z_\D)$ denotes integration
with respect to the $z_\D$ variables introduced in Eq.\equ{e2.1}.

The plan is then to integrate the $z_\D$ variables for $\D$ on a given
scale and prodeed to integrate the other $z$-variables ``one scale at a
time'': the correct question to pose is whether the parameters $\l,\m,\n$
can be so chosen {\it as functions of $N$} in such a way that the limit as
$N\to\infty$, called {\it ultraviolet limit}, of
\be \kern-3mm S_N(x_1,\ldots x_{2s})\defi\kern-2mm 
\int \kern-1mm \f_{x_1}\f_{x_2}\cdots\f_{x_{2s}}\,
\frac{e^{-\int_\L(\l(\f_x^{[\le N]})^4+\m ( \f^{[\le N]}_x)^2+\n)\,dx}
  P(d\f)}{Z_N}
\Eq{e2.5}\ee
is not only well defined for all pairwise distinct $x_1,\ldots ,x_{2s}$ and
all $s$.  but it is also ``non trivial'', \ie it is not computable via
Wick's rule from $S_\infty(x_1,x_2)$ (which means that after removing the
cut-off, $N\to\infty$, the theory is not a free theory).

In applications the physically relevant quantities are expressed in terms
of the {\it Schwinger functions}, $S_\infty(x_1,\ldots x_{2s})$: so on the
one hand the bare constants disappear and, on the other hand, one is left
with the problem of checking that the $S_\infty(x_1,\ldots x_{2s})$ have
the properties needed to describe a theory that agrees with the basic laws
of dynamics: which essentially amount at suitable analyticity properties of
the Schwinger functions, \cite{OS973}.

The point of the hierarchical model is that the construction of its
Schwin\-ger functions as limits of regularized probability distributions of
the fields $\f_x$ presents the same difficulties, {\it in dimension $2$ and
  $3$}, that are encountered in the study of the integrals like
Eq.\equ{e1.3}. 

Namely attempting an expansion in powers of the couplings leads to
divergent quantities which can be eliminated through suitable
resummations. Its study via Wilson's renormalization group method {\it
  simply avoids introducing divergences}.

\def\SEC{Effective potentials and running couplings}
\section{\SEC}\iniz\label{sec3}

A first key remark is that if in the integral Eq.\equ{e2.4} the integration
is performed only with respect to the $z_\D$ with $\D\in \QQ_N$ then the
computation can be performed via perturbation theory and {\it with complete
  control of the remainders}. The argument of the exponential should be
appropriately regarded as a function of the ``ultraviolet $z_\D$'s''; let
for $\D\in\QQ_N$
\be X^{[\le N]}_\D\defi \frac{\f^{[\le N]}_\D}{\sqrt\media{(\f^{[\le
        N]}_\D)^2}} =\a_N z_\D+\b_N X^{[< N]}_{\D'} \Eq{e3.1}\ee
where $\D\subset \D'\in\QQ_{N-1}$, and $\a_N^2=\frac{2^{(d-2)N}}
   {\sum_{k=0}^N 2^{(d-2)k}}$, $\b_N^2=1-\a_N^2$, so that
 
\be\eqalign{\a^2_N=&\frac1{{N+1}},\ \ \ \, \b^2_N=\frac{N}{{N+1}},
\qquad{\rm if}\quad d=2\cr
\a^2_N,&\b_N^2=\frac12+O(2^{-(d-2)N}), \qquad{\rm if}\quad d=3\cr}
\Eq{e3.2}\ee
In the following the $O(2^{-(d-2)N})$ will be neglected (for the purpose of
simplified notations). 

Since the volume of $\D$ is $2^{-dN}$ the integrals in the exponential are
\be
\eqalign{
\LL(X^{[\le N]})=&
\sum_{\D\in\QQ_N} (\l 2^{-dN} C_N^2  (\a_N z_\D+\b_N X^{[<N]}_{\D'})^4 
\cr
&\kern1.cm+\m 2^{-dN} C_N (\a_N z_\D+\b_N X^{[<N]}_{\D'})^2 +2^{-dN}\n)\cr
=&\sum_{\D\in\QQ_N} V_N({\a_Nz_{\D}+\b_N X_{\D}})\cr}
\Eq{e3.3}\ee
where $C_N\defi\media{(\f^{[\le N]}_\D)^2}$, \ie $C_N=1+N$ if $d=2$ and 
in general $2^{(d-2)N}(1+O(2^{-(d-2)N}))$  if $d>2$.

Therefore in performing the integral over $z_\D$ the variable $z_\D$
appears multiplied by a factor $ 2^{-d N}C_N^2\sim 2^{-(4-d)N}$ or
$2^{-dN}C_N\sim  2^{-2N}$.\ \footnote{Here $\sim$ means that the equalities
are true in dimension $d=2$ up to a factor $N^2$ or $N$ or up to a factor
$O(1+2^{-(d-2)N})$ in dimension $d=3$.}

For definiteness suppose hereafter that $d=3$ (the case $d=2$ is actually
much simpler) and, to simplify notations, take $\a^2_k,\b^2_k$ to be
$\a_k^2=\b_k^2=\frac12$ (thus neglecting the mentioned correction of
$O(2^{-N})$). 

Call $\l_N,\m_N,\n_N$ the ``bare coupling constants'' in $\LL_N$: the {\it
  ultraviolet stability problem} is to show that the couplings can be
determined so that the $Z_N$, Eq.\equ{e2.4}, as well as all Schwinger
functions, Eq.\ref{e2.5}, are bounded above and below uniformly in $N$ and
{\it cannot be evaluated by a Wick rule starting from $S(x_1,x_2)$}.

The idea is to define the ``effective potential'' $V_k$ on scale $k< N$
as
\be e^{\sum_{\D'\in\QQ_{k}} V_k(X_\D')}=
\int  \prod_{\D'\in\QQ_{k}}\Big(
\prod_{\D\subset\D'} e^{V_{k+1}(\frac{z_\D+X_{\D'}}{\sqrt2})}
\frac{e^{-\frac12 z_\D^2}}{\sqrt{2\p}}dz_\D\Big)
\Eq{e3.4}\ee
The hierarchical structure reduces the study to the recursion
\be e^{V'(X)}=\Big(\int  
e^{V(\frac{X+z}{\sqrt2})} P(d z)\Big)^{2^3}
\Eq{e3.5}\ee
and it has to be shown that starting with a polynomial of degree $4$ in
$X$, of the form $V_N(X)=\l_{0,N}+\l_{1,N}:X^2:+\l_{2,N} :X^4:$, and {\it
  fixed $p>3$} the recursion defines a sequence of effective potentials
$V_k(X)$ which, up to a remainder $\h_k=O(\l^p 2^{-(p-3)k})$ with
$\l\defi\l_{2,N}$, is a polynomial $\LL_k(X)$ of degree $2p$:%
\footnote{\small Rather than in terms of the monomials $X^n$ it will be
  expressed in terms of Wick's monomials $:X^n:$, because this simplifies
  the algebra (if the calculation of several needed Gaussian integrals is
  performed via Feynman's graphs, reducing substantially their
  number). Recall that Wick's monomials of a Gaussian variable $X$ are
  defined in terms of the Hermite polynomials $H_n(X)$ (with leading
  coefficient $2^{n}$) as $:X^n:\defi\Big(\frac{C}2\Big)^{\frac{n}2}
  H_n(\frac{X}{\sqrt{2C}})$, with $C=\media{X^2}$, \cite[8.950.2]{GR965}.}
\be\LL_k(X)=-\sum_{n=0}^{p} \l_{k,n} :X^{2n}:\Eq{e3.6}\ee
and $\l_{k,n}$ are called {\it running couplings} on scale $k$.

In other words the effective potential $V_k$ on scale $k$ is a polynomial
of degree $2p$ within a remainder, of order $\l^p$, {\it summable over $k$
uniformly in $N$}. 

The recursion is therefore reduced to a polynomial map in
$p$ dimensions, if the analysis has to be performed up to a remainder
$\l^p$. In the present work the theory of the recursion, {\it \ie of the
  beta function}, will be presented and reduced to the iteration of a map
involving finitely many ``running couplings'' in dimenson $d=3$: a point of
view which was not literally followed in the earlier works on the
hierarchical model, \cite{Ga978,BCGNOPS978}. \*

\0{\it Remarks:} (1) The ultraviolet problem is essentially
reduced to prove that the ``trivial fixed point'', $V=0$, of the recursion
Eq.\equ{e3.5} is unstable and therefore, if after $N$ iterations a
$N$--independent non trivial result is desired, it is possible to start with
a $V$ close enough to $0$ so that after the $N$ steps it evolves into a
$O(1)$ final $V_0$. 
\\
(2) In other words in the ultraviolet problem the ``bare
couplings'' tend to $0$ as the cut-off $N\to\infty$ and the problem can be
studied via perturbation theory {\it if the large values of the fields can
  be controlled} (note that no matter how small is $\l_N$ there will always
be fields so large that $V$ is large). 
\\
(3) The infrared problem, directly related to the critical point
theory, cannot be studied by simply reducing it to the analysis of a
polynomial map.  Since the recursion is the same in the ultraviolet and
infrared problems, what makes the analysis easy in the ultraviolet problem
makes it difficult in the infrared problem, where the role of the trivial
fixed point has to be played by another fixed point $V^*$ which is non
trivial and unstable so that by starting close enough to it it is possible
to stay close to it until the infrared cut-off is reached.
\\
(4) Wilson used a computer aided approach to show the existence of the non
trivial fixed point in dimension $d=2,3$. This was an important result also
because it made clear, in a concrete case, that the idea of the fixed point
was a generalization of the Gell-Mann-Low eigenvalue condition for the bare
coupling constant of quantum electrodynamics, \cite{Wi970}, and opened the
way to the understanding of the critical point scaling properties.  A
rigorous determination of the existence and of several analytic properties
of $V^*$ have been later studied in the remarkable works
\cite{KW986,KW991}.

\def\SEC{The beta function}
\section{\SEC}\iniz\label{sec4}

In superrenormalizable theories, like $\f^4$ in dimension ($2$ or) $3$, the
beta function is a polynomial transformation mapping the coupling constants
on a scale $k+1$ into the couplings on scale $k$. Its definition is based
on the formal integration with respect to the Gaussian
$P(dz)\defi\frac{e^{-\fra12 z^2} dz}{\sqrt{2\p}}$

\be (\int e^{\LL(\frac{X+z}{\sqrt2})} P(dz))^{2^3}
=\exp 2^3\sum_{n=0}^\infty\frac1{n!}
\media{\LL(\frac{X+z}{\sqrt2})^n}{}^T\Eq{e4.1}\ee 
where ${}^T$ indicates that the $\media{\LL^n}$ is the order $n$ truncated
expectation with respect to the Gaussian variable $z$.\footnote{\small The
  $n$-th truncated expectation of a random variable $Y$, {\it with any
    distribution}, also called the $n$-th ``cumulant'', is defined as
  $\media{Y^n}^T =\dpr^n_\e \log \media{\e Y}\Big|_{\e=0}$.}

The heuristic reason behind the procedure is in the comment following
Eq.\equ{e3.3}: once reduced the field $\f^{[\le N]}_x$ to
$2^{\frac{d-2}2N}X^{[\le N]}_x$, \ie to a quantity of order $1$ times its
(average) size $2^{\frac{d-2}2N}$ and after extracting the size $2^{-dN}$ of
the volume element over which the field of scale $\le N$ is constant, it
remains to integrate over $z_\D$ the exponential of a sum of {\it very
  small} quantities, of $O(\l 2^{(d-4)N})$, functions of the $z_\D$; 
therefore it looks possible (and even apparently easy) to use explicit
perturbation methods (\ie evaluate the integrals via Taylor's expansions).

A perturbation method \footnote{\small Usually called in this context
  ``exact'' as it is not merely a formal expansion but provides exact
  results once the tolerance of the approximation is, arbitrarily, prefixed
  and if the physical couplings of the theory are small enough (but neither
  infinitesimal nor of size depending on the approximation order $p-1$).}
will stop at some order and the remainder will have to be carefully
estimated. It is clear that the best that it is possible to hope is that if
perturbation calculations are pushed to order $p-1$ the remainder will be
at least of the $p$-th power of the small parameter, \ie $O((\l
2^{(d-4)N})^p)$.

The error will be repeated once per each of the $2^{dN}$ boxes $\D\in
\QQ_N$ and this will add up to $O((\l 2^{(d-4)N})^p\,2^{dN})$: {\it
  therefore the calculation of the integral has to be performed up to order
  $p-1$ such that $(d-4)p+d<0$ which means $p\ge2$ if $d=2$,
\ie a calculation to first order is sufficient (which makes the problem a
bit too easy), and $p\ge 4$ if
  $d=3$}: where an exact calculation is necessary at least to order $3$.

Of course after the first integration the effective potential on scale
$N-1$ will be quite different from the initial $\LL(X)$: therefore the
parameters initially in $\LL$ will have to be adjusted so that the form of
the new $\LL'$ is as close as possible to that of $\LL$ and the procedure
can be iterated. 

This puts a severe constraint on the initial parameters:
it imposes that upon integration they change according to a precise rule,
called the {\it beta function} constraint.

Let $\LL(X)$ be a polynomial of degree $2p$ as in Eq.\equ{e3.6}.
Given $p$ the beta function is obtained by replacing the {\it r.h.s.}
series in Eq.\equ{e4.1} (which at best is asymptotic) by its
``approximation''

\be \LL'(X)=2^3\Big(\sum_{n=0}^{p-1} \frac1{n!}
\media{\LL(\frac{X+z}{\sqrt2})^n}{}^T_p\Big)\Eq{e4.2}\ee
where the $\LL'$ in the \rhs is calculated by 
\*

\0(1) first compute the truncated expectations $\wt\LL'(X)=2^3\sum_{n=1}^{p-1}
\media{\LL(\frac{X+z}{\sqrt2})^n}{}^T$, for instance using Wick's rule.
The result will be a polynomial in the constants $\l_k,\,k\ne2$ in
$\LL(X)$, see Eq.\equ{e3.6}, with coefficients depending on $X$.
\\
(2) assign degree $1$ to the coefficient\footnote{\small As the $\f^4$
  model is being studied.} $\l_2$ of $:X^4:$ and degree $\ge2$ to the other
constants $\l_k,\,k\ne2$ and then truncate the polynomials in the $\l_k$ by
retaining only their monomials of degree $<p$.  \\
(3) Express the even polynomial of degree $2p$, thus obtained, again on the
Wick's monomials basis and call it $\LL'(X)$: it will have the form
Eq.\equ{e3.6} with suitable coefficients $\l'_k$.  \*

Therefore the transformation $\LL\to\LL'$ maps $\{\l_n\}_{n<p}$ into
$\{\l'_n\}_{n<p}$. For instance:
\be \eqalign{p=1 \to &\ \LL'(X)=0, \qquad\cr p=2 \to &\ \LL'(X)=2\l_2:X^4:\cr 
p=3\to
  &\ \LL'(X)=2^3\l_0+2^2\l_1 :X^2: +2\l_2:X^4:+\l_3 :X^6:\cr
&+\l_2^2(a_6 :X^6:+a_4
  :X^4:+ a_2 :X^2:+a_0)\cr}\Eq{e4.3}\ee
and for $p=4$, calling $\l_0\equiv
\n,\l_1\equiv\m,\l_2=\l,\l_3\equiv\s,\l_4\equiv\th$
\be\eqalign{
\n'=&2^3\n+a_0\l^2+(d_0 \l^3)\cr
\m'=&2^2\m+a_2\l^2+(b_2\l\m+c_2\l^3+d_2\l\s)\cr
\l'=&2\l+a_4\l^2 + (b_4\l\m+ c_4\l^3+d_4 \l\s+e_4\l\th)\cr
\s'=& \s+ a_6\l^2+(c_6\l^3+d_6\l\s+e_6\l\th)\cr
\th'=&2^{-1}\th +(a_8\l^3+d_8\l\s+e_8\l\th)\cr}\Eq{e4.4}\ee
The first three constants are called {\it relevant couplings}, the fourth
is called {\it marginal} and the fifth {\it irrelevant}.  The coefficients
$a_j,b_j,c_j,d_j,e_j$ can be computed exactly via elementary integrations:
they have a combinatorial nature and are expressible in terms of Feynman
graphs.
 
Needless to say the qualification ``irrelevant'' is not supposed to convey
an implication of ``negligible''; on the contrary the irrelevant terms are
very important and needed in the applications of the theory. The whole
problem is to control them and their contributions to the remainders.  For
larger $p$ similar relations hold and more ``irrelevant'' terms arise.

From now on $p=4$ will be fixed, once understood this case it should be
clear how to treat the cases $p>4$ and no new problems will arise: {\it by the
above comments} (about the errors, see the two paragraphs preceding
Eq.\equ{e4.2}) {\it this is the lowest possible choice of $p$}.  

The Eq.\equ{e4.4} maps $(\l,\m,\n,\s,\th) \to(\l',\m',\n',\s',\th')$: since
the origin is an unstable fixed point (in three directions and marginal in
one) there will be a trajectory which starting close to $0$ in $N$ steps
reaches a point at finite distance from the origin; one checks (by
substitution) that for $k=0,\ldots, N$:
\be\eqalign{
\l_k=&\l 2^{-k},\qquad \m_k=-2^{-2k} k a_2\l^2,\qquad
\n_k=2^{-2k-1}\,a_2\,\l^2\cr
\s_k=&2^{-2k}\l^2 s_{6,N},\qquad
\th_k=2^{-3k}\l^3 t_{8,N}\cr
}\Eq{e4.5}
\ee
with $s_{6,N}=a_6\sum_{n=1}^{N-1} 2^{-2n}$, $t_{8,N}=d_8
\sum_{n=1}^{N-1}\frac{2^{-3n}}{2^n}$ and $a_2,a_6,d_8$ suitably chosen, is
a trajectory of the map for $k=0,\ldots,N-1$ if $\s_N=\th_N=0$ {\it up to
  corrections amounting at factors $(1+ const\, k 2^{-k})$ in each term}:
for instance a correction to $\n_k$ is $-2^{-3k}\,k\, d_0\,\l^3$, for a
suitable (precise) choice of $d_0$ and to $\s_k$ a correction is
$\sum_{n=1}^{N-1} d_6 \l^3 2^{-3(k+n)}$, and there are other similar
corrections to the trajectory in Eq.\equ{e4.5}; here empty sums mean $0$.

In the next section it will be shown that the existence of a trajectory
with the properties Eq.\equ{e4.5} with $\l>0$ (a quite elementary fact) is
all what is needed for a complete analysis.

\def\SEC{The renormalization group}
\section{\SEC}\iniz\label{sec5}

Given a polynomial $\LL(X)$ with the property that there are constants
$m>0$ and $B>B'>\lis B$ such that $B>\frac{B'\pm\lis B}{\sqrt2}>\frac12B$ 
(\eg $B'=B(1-\frac18), \lis B=\frac B8 $), $B>1$, and
 
\be\eqalign{
\LL(X)&<0,\quad B>|X|>\frac{B}2,\quad{\rm and}\cr
\LL(X)&<m, \quad |X|< B\,.\cr}
\Eq{e5.1}\ee
A concrete case to keep in mind could be $\l:X^4:+\m :X^2:+\n$ with
$:X^{2k}:= 2^{-\frac{k}2}H_{2k}(\frac {X}{\sqrt2})$ with $\l>|\m|,|\n|$ and
$B$ large enough.

Then, for $Y\defi\frac{X+z}{\sqrt{2}}$ and $\ch(condition)\defi1$ if conditon
  is true, $\defi0$ otherwise:
\be\eqalign{ 
\int&  e^{\LL(Y)}P(dz)\le \int e^{\LL(Y)\ch(|Y|<B)} dP\cr
&\le\ch(|X|>B') \int e^{\LL(\frac{X+z}{\sqrt{2}})\ch(|Y|<B)}
\Big(\ch(|z|>\lis B)
+\ch(|z|<\lis B)\Big)dP
\cr
&\kern3mm 
+\ch(|X|<B') \int e^{\LL(\frac{X+z}{\sqrt{2}})\ch(|Y|<B)}\Big(\ch(|z|>\lis B)
+\ch(|z|<\lis B)\Big)dP
\cr}\Eq{e5.2}\ee
If $||\LL||=\max_{|Y|< B} |\LL(Y)|$ then (making use of
$|\frac{X+z}{\sqrt{2}}|\ge \frac{B'-\lis B}{\sqrt2}>\frac12B$ for $|X|\ge
B',|z|<\lis B$, of Eq.\equ{e5.1} and of Taylor's remainder estimate)
\be\eqalign{
&\ch(|X|>B') \Big(e^m e^{-\frac12 \lis B^2}+ 1\Big)
\cr
&\ch(|X|<B') \Big(e^m e^{-\frac12 \lis B^2}
+\exp\Big(\sum_{n=1}^{p-1}
\frac{\media{\ch\LL^n }{}^T}{n!}+c'_p||\LL||^p
\Big)\Big)\cr}
\Eq{e5.3}\ee
where $\ch\equiv\ch(|z|< \lis B)$ and $c'_p$ is a constant depending only
on $p$. Hence $|\media{\ch\LL^n }{}^T-\media{\LL^n }{}^T|\le ||\LL||^n
e^{-\frac{\lis B^2}2}$, there is $c_p$ such that
\be\eqalign{
& \sum_{n=1}^{p-1}\frac{\media{\ch\LL^n }{}^T}{n!}
\le \sum_{n=1}^{p-1}\frac{\media{\LL^n }{}^T}{n!} +
e^{-\frac12\lis B^2}\sum_{n=1}^{p-1}\frac{||\LL||^n}{n!}\cr
&|\media{\LL^n }{}^T-\media{\LL^n }{}^T_p|\le \L^p B^{2p} c_p,
\qquad\L=\max_{0\le n\le 2p} |\l_n| \cr}\Eq{e5.4}
\ee
Therefore
\be \int e^{\LL(Y)\ch(|Y|<B)}dP
\le (1+e^m e^{-\frac12\lis B^2})^{2^3}e^{\LL'(X)\ch(|X|<B')}\Eq{e5.5}\ee
Introduce sequences $B_k,\lis B_k$ are such that for all $k\ge0$
\be B_k= (k+2)^4b,\quad B'_k= B_{k-1},\quad
\lis B_k= (k+2)^2b\Eq{e5.6}\ee
for a constant $b>0$ to be fixed later (as $b=12$).

Let
$1>\l_k>0,\m_k,\n_k,\s_k,\th_k$ be a trajectory of the flow generated by
the beta function with $|\m_k|,|\n_k|,|\s_k|,|\th_k|<\l_k^2 (1+k)$ 
satisfying Eq.\equ{e4.5}. 
Notice that

\be \eqalign{
:X^4:\equiv& X^4-6X^2+3\ge-6,\ {\rm and}  \cr
:X^4:>&\frac12 X^4,\ |X|\ge12=b\cr}\Eq{e5.7}\ee
therefore if $2^{-k}\l> 2^{-2k}k \l^2 c$ for suitable $c,m$ it will be

\be\eqalign{
 \LL_k(X) <&0, \qquad {\rm for}\quad  B_k>|X|>\frac{B_k}2,\cr
 \LL_k(X)<& m  \qquad {\rm for}\quad  |X|<B_k,\cr}
\Eq{e5.8}\ee
and Eq.\equ{e5.1} hold with $B'=B_{k-1}$. It follows

\be \eqalign{
V_k(X)\le &\sum_{\D'\in\QQ_k} \sum_{\D\subset \D'}
\Big(\sum_{n=1}^{p-1} \frac1{n!}
2^3\media{\LL(\frac{X_{\D'}+z_\D}{\sqrt2})}{}^T_p\Big)\cr
&+\sum_{j=k}^N 2^{3(j+1)}\log(1+e^m e^{-\frac12 \lis B_{j}^2})
\defi V^0_k(X)+\e^+_k\cr}\Eq{e5.9}\ee
and $\e^+_k\le \e^+_0\defi \sum_{k=0}^\infty 2^{3(k+1)}\log(1+e^m
e^{-\frac12 \lis B_{k}^2})$ is an estimate of the total error on $V_k$ for all
$j$.

In other words the value of $\log Z$ is determined via an asymptotic
expansion with finite coefficients, {\it provided} a lower bound coinciding with
the upper bound up to order $p-1$ and with an error estimate of the same
size as that on the upper bound. 

A lower bound can be easily constructed simply by restricting the
integration domain:
\be \int e^{-\LL_N(X)}P(dz)\ge \int e^{-\LL_N(X)}\prod_{k=0}^N\prod_{\D\in
  \QQ_k} \ch(|z_\D|<\lis B_{k})\, g(dz_\D)
\Eq{e5.10}\ee
Since $|z_\D|<\lis B_k$, for $\D\in\QQ_k, \forall k$ implies $|X_\D|<B_k$
it appears that the estimate is essentially the same as the one used to
find the upper bound to the last of the integrals in Eq.\equ{e5.2}: 
the result is similar to Eq.\equ{e5.9}  the integral yields 
\be V_k(X)\ge V^0_k(X)-\e^-_k\Eq{e5.11}\ee
where $\e^-_k=-\sum_{j=k}^N 2^{3(j+1)}\log(1-e^m e^{-\frac12 \lis
  B_{j}^2})\le\e^-_0$.

This is iterated leading to a lower bound $e^{\n_0-\sum_{k=0}^\infty \e^-_k
  2^{3k}|\L|}$, proceeding as in the upper bound.

 Finally the errors $\e^\pm_k$ sum up to a quantity that is $o(\l^3)$
 provided the constants $B_k,\lis B_k$ have the form $B(\l) (k+1)^a, \lis
 B(\l) (k+1)^b$ and $B(\l)$ is chosen so large that the error due to the
 truncation of the $z$ integrals which contain $e^{-\frac12 \lis B^2_{k}}$
 become more infinitesimal than any power (hence not affecting corrections
 of any order in $\l$): this can be achieved simply by $B(\l)=B
 (\log(1+\frac1\l))^2$ and $B>1$, \cite{Ga978, BCGNOPS978}.
\*

\0{\it Remarks:} {\bf(1)}: \rm It must be stressed that the possibility of the
  iteration with controlled remainders relies on the possibility of
  eliminating the ``{\it large fields'}'' at the first integration (\ie on
  scale $N$) and replacing $\LL_N$ with $\LL_N \ch$ controlling the error:
  which could only be done because $\s_N,\th_N=0$; as a consequence they
  will never grow enough to affect the positivity of $\LL_k$ which remains
  controlled by $:X^4:$ as long as $|X|$ is bounded by a power of $k$,
  because the coefficients of the other terms of $\LL$ will be
  exponentially small relative to the coefficient of $:X^4:$.
\\
{\bf(2)} Analysing the proof it is seen that the $\LL(X)$ could have been
kept a polynomial of degree $4$: namely $\LL(X)=\n_N+\m_N :X^2:+\l_N :X^4:$
defining the beta function by Eq.\equ{e4.4} with $\s,\th=0$. The upper and
lower bounds would have been obtained in the same way (including the
contributions with $\s,\th$ in the error). The procedure followed has been
chosen because it can be extended to all $p\ge4$ to prove that the
perturbatiion theory yields upper and lower bounds correct to any prefixed
order. It can also be extended to obtain bounds on the Schwinger functions.
\\
{\bf(3)} A natural question is whether the $d=4$ case can be studied in a
similar way. In this case $d-4=0$ and the only small parameter can be found
among the bare couplings. No power of $2^{-N}$ helps, thus spoiling the
main tool which consisted in taking advantage of the $2^{(d-4)N}$
dimesionless size of the interaction coupling. Nevertheless a formal theory
of the resummation is possible, see \cite{Ga985b} for a beta function
analysis, in the case of $\f^4$ model {\it on $R^4$}: but {\it not in the
hierarchical case}. The hierarchical case could be studied if the recursion
\be e^{ V'(X)}=\Big(\int e^{V(\sqrt{\frac{3}{4}}z+\sqrt{\frac14}X)}
P(dz)\Big)^{2^4}\Eq{e5.12}\ee
which is the $d=4$ version of the $d=3$ Eq.\equ{e3.5}, had an unstable
fixed point. However, {\it as Wilson pointed out}, \cite[endnote 8]{Wi972},
no such fixed point could be found, neither by theoretical investigations
nor by computer assisted search. The latter all indicate that, on the
contrary, no matter which choice of the bare couplings was made the only
possiblity for the final Schwinger functions would be that they were the
free field functions.  
\\
{\bf (4)} The non hierarchical case is very different but, although a
formal resummation is possible the beta function that drives it can only be
defined as a formal power series.  In spite of several results supporting
the conjecture that it is impossible to obtain obtain nontrivial Schwinger
functions in a scalar quantum field theory in dimension $4$ is still (wide)
open, \cite[endnote8]{Wi972},\cite{GR984}.
\\
{\bf (5)} The models $\f^6$ in $d=3$ is only superficially similar to the
$\f^4$ in $d=4$: in the hierarchical case it still appears to lead to a
trivial result or possibly, if $\l_N=\l 2^{-\frac{N}2}$, back to the $\f^4$
case. However in dimension $3$ it was a major discovery by Wilson,
\cite{Wi972}, that (in the hierarchical case) it admits a non trivial
theory different from the $\f^4$ one: \ie a non trivial fixed point $V^*$
which is unstable in only one direction (in the space of the $V$'s). Its
stable manifold is crossed by the family of $V$'s of the form $r X^2+\l
X^6$ as $r$ varies reaching a critical value $r_c(\l_0)$. Therefore the
stable manifold of $V^*$ can play the same role of the trivial fixed point
for the $\f^4$ model discussed above. Starting $V_N=r_N X^2-\l_0 X^6$ with
$r_N$ close enough to the critical $r_c(\l_0)$ the $V_k$ are exponentially
repelled by the stable manifold of $V^*$ and reach a finite distance from
$V^*$ on scale $1$. The $V^*$ can also be used to obtain a nontrivial
infrared behavior: if $r=r_c(\l_0)$ the $V_k$ for $k<0$ will approach
$V^*$, and a scale invariant long distance family of Schwinger functions
describing a critical point of a model in which $r-r_c(\l_0)$ plays the
role of $T-T_c$. Changing $\l_0$ (Wilson fixes $\l_0=0.1$) only changes the
critical value $r_c(\l_0)$ and has no influence on $V^*$. A rigorous proof
of the existence of $V^*$ in dimensions $2,3$ is, as mentioned above, in
\cite{KW986,KW991}.
\\
{\bf(6)} In dimension $d=2$ it is possible with the renormalization group method
(whether hierarchical, very easy, or in the non hierarchical model) to
check that $\f^{2n}$ can be defined for all $n$: this was the first case in
which ultraviolet stability was established, \cite{Ne966}, via an
alternative approach that, however, could not be extended to $d=3$, not
even in the $\f^4$ model. In dimension $3$ only the $\f^4$ can be treated,
essentially along the lines of the above hierarchical analysis.

\def\SEC{References}  
\bibliographystyle{plain}

\end{document}